\documentclass{article}

\usepackage[pdftex]{graphicx}

\usepackage[cmex10]{amsmath}
\usepackage{array}
\usepackage{mdwmath}
\usepackage{mdwtab}
\usepackage{eqparbox}
\usepackage{url}

\begin{document}
	
\title{Electrostatic precipitation of exhaled particles for tensiometric examination of pulmonary surfactant}

\author{Andrey~Shmyrov, Alexey~Mizev, Irina~Mizeva and Anastasia~Shmyrova}

\thanks{This study was supported Russian Foundation for Basic Research under project No. 17-41-590095. }
\thanks{A. Shmyrov, A. Mizev, I. Mizeva and A. Shmyrova are with the Institute of Continuous Media Mechanics, Perm, Russia  (correspondence e-mail: mizeva@icmm.ru).}

\maketitle

\begin{abstract}
\textbf{\textit{Objective:}  Collecting exhaled particles that represent small droplets of the alveolar lining fluid  shows great promise as a tool for pulmonary surfactant (PS) sampling. For this purpose, we present a setup consisting of two modules, namely, a module for droplet collecting, called an electrostatic aerosol trapping (ESAT) system, and a measurement module for studying PS properties. We suggest the way how to extract numerical values from the experimental data associated with PS properties.}

\textbf{\textit{Methods:} The operating principle of ESAT is based on the electrostatic precipitation of exhaled particles. The native material was collected directly on the water surface, where an accumulated adsorbed film of PS was examined with tensiometric method. The modified capillary waves method adapted to study small volume samples was utilized. The efficiency of the setup was verified in the experiments with healthy subjects.}

\textbf{\textit{Conclusion:} The accumulation of PS components on the water surface in an amount sufficient for tensiometric study was reported. It was shown how to extract the numerical values from the experimental data characterizing  PS properties.} 

\textbf{\textit{Significance:} The idea underlying the new concept used in this study may give impetus to further development of point-of-care facilities for collecting PS samples and for their express analysis.}  
\end{abstract}

% === KEYWORDS =======================================================
%\begin{IEEEkeywords}
%Exhaled particles, pulmanory surfactant, surface tension,  alveolar lining %fluid, capillary waves, point-of-care testing 
%\end{IEEEkeywords}

%\IEEEpeerreviewmaketitle

\section{Introduction}
\label{sec:intro}

Human breathing is accompanied by the formation of small droplets of the alveolar lining fluid (ALF), which are emitted from the lungs with exhaled air in aerosol form. The droplet formation mechanism is associated with closure and reopening of airways during breathing \cite{Haslbeck2010,Johnson2009,almstrand2010effect}. At the end of each exhalation the thickness of the fluid layer, lining bronchioles, becomes comparable with the airway radii. This leads to the development of the  Rayleigh instability \cite{Grotberg2011} which, in turn, results in the formation of a liquid plug blocking the airway. This phenomenon, known as airway closure, occurs in normal individuals near the end of expiration and, as a rule, takes place only in the thin airways located in the distal lung regions \cite{Holmgren2010}. During the next inspiration the plug ruptures and the airway reopens, which is accompanied by droplet formation. The exhaled air of a healthy human during normal breathing contains, on average, a few submicron particles per cubic centimeter (\cite{Fairchild1987,Papineni1997,morawska2009size}. Although both the size distribution and the particle concentration demonstrate high reproducibility within one subject, there are also a high inter-subject variability \cite{Schwarz2010} and a strong dependence on the type of breathing and the breathing maneuvers before sampling \cite{johnson2011modality, bake2017exhaled}. Analysis of the droplet composition has revealed that the aerosol particles originate from ALF and contain all its components in undiluted concentration \cite{Olin2013, Tinglev2016, Ullah2015}. Significant differences were established between the aerosol characteristics and droplet compositions collected from the groups of subjects with chronic obstructive pulmonary disease \cite{schwarz2015characterization}, asthma \cite{schwarz2015characterization}, cystic fibrosis \cite{Almstrand2009} or pulmonary tuberculosis \cite{Wurie2016} and healthy subjects. 

The emitted droplets are the ALF microsamples, and therefore their trapping provides a non-invasive way for obtaining the native material directly from the respiratory tract. This can be an alternative to traditional lung sampling methods such as the bronchoscopy procedure, which is invasive and uncomfortable. A certain amount of non-volatile ALF components can be detected in the exhaled breath condensate (EBC) \cite{horvath2003exhaled}. Since this method was specifically developed for the volatile components collection, the capture of exhaled droplets during condensation of exhaled air occurs, for the most part, accidentally. This leads to extremely small capturing efficiency and rather low repeatability of the results \cite{kazakov2000dynamic}. Modern technologies for sampling aerosol particles employ different collection mechanisms, including impaction, impingement, filtration, and electrostatic precipitation \cite{baron2001aerosol}. As applied to airborne bioaerosols, these techniques are commonly utilized for capturing biological agents such as pathogens, microorganisms or viruses, from ambient air. Another important feature of the exhaled particles is that they contain pulmonary surfactant (PS). In this work, we primarily focus on understanding and evaluating the opportunity to collect PS for further examination. 

PS is the main component of the ALF and a surface-active lipoprotein complex produced in a human lung by type II alveolar cells \cite{notter2000lung}. The most important function of PS is to reduce the alveolar surface tension, which results in increasing pulmonary compliance and allows the lung to inflate much more easily, reducing thereby the work of breathing. Moreover, the unique ability of the compressed PS layer to decrease surface tension to a very low, almost zero, level also prevents atelectasis (collapse of lung alveoli) at the end of expiration. Each component of the lipoprotein complex plays a role in the formation of surface properties, and any change in the composition leads to a reduction in the surface activity of the PS and, as a result, to dysfunction of the entire pulmonary surfactant system \cite{longo1993function,schurch2001surface,Possmayer2001,ma2006real,bykov2019dynamic}. A variety of pulmonary diseases (such as asthma, pneumonia, adult respiratory distress syndrome, tuberculosis, etc.) are able to cause surfactant deficiency or to change its composition, which reduces the PS surface activity, provoking alveolar instability and development of inflammatory processes in the lung \cite{Hohlfeld2002, Baritussio2004, Wright2001, Schwab2009, Raghavendran2011, Chroneos2009, Chimote2005, Hasegawa2003, Wang2008}. From this point of view, the study of the surface-active properties of PS, collected in the course of aerosol particles capturing, is an effective way to noninvasively monitor the functional status of a lung surfactant system. 

In this paper, we present the setup which combines two modules, namely, a module for droplet collecting, called an electrostatic aerosol trapping (ESAT) system, and a measurement module for studying PS properties. The operating principle of ESAT is based on electrostatic precipitation of exhaled particles on the water surface. The concept was first proposed in \cite{PARDON2015}, and the developed system was tested on model aerosol, yet it was not applied to human bioaerosol analysis. In \cite{Morozov2017}, a collection system for dry solid residues from exhaled breath was introduced and the material collected at a solid surface was studied via atomic force microscopy. The key advantage of the capturing method described here is that it allows gathering the native material in a quantity sufficient for tensiometric examination. The PS containing in droplets adsorbs on the water surface and changes its surface properties. The measurement module of our setup utilizes the modified capillary waves method \cite{Shmyrov2019} which is very sensitive to changes in surface properties. 

The paper is structured as follows. Section ~\ref{sec:ESAT} includes the design and operation principle of the ESAT system. Section ~\ref{sec:ESAToptimization} contains a description of the experiments with a model aerosol performed to determine an optimal configuration of the ESAT and to find the most efficient trapping regime. The design of the module for the droplet collecting, including the exhaled air
delivery system and the ESAT system, is presented in section \ref{sec:Exhaled air} in detail. Section \ref{sec:capwave} provides a brief description of the modified capillary waves technique, which has been previously described in detail in \cite{Shmyrov2019}. Section~\ref{sec:dataprocessing} discusses the data processing procedure for calculating the wavelength and attenuation coefficient of the capillary wave from the acquired initial interferograms. Section~\ref{sec:res} presents the data illustrating the efficiency of the setup and its components, collecting and measuring modules, for examination of the aerosol produced during the exhalations of a healthy subject (sec.~\ref{sec:res}). Section ~\ref{sec:concl} summarizes the obtained results and describes possible applications of the proposed technique.

\section{Experimental setup and methods}
\subsection{Design and operating principle of the electrostatic aerosol trapping (ESAT) system}
\label{sec:ESAT} 

\begin{figure}[h]
\includegraphics[width=0.5\linewidth]{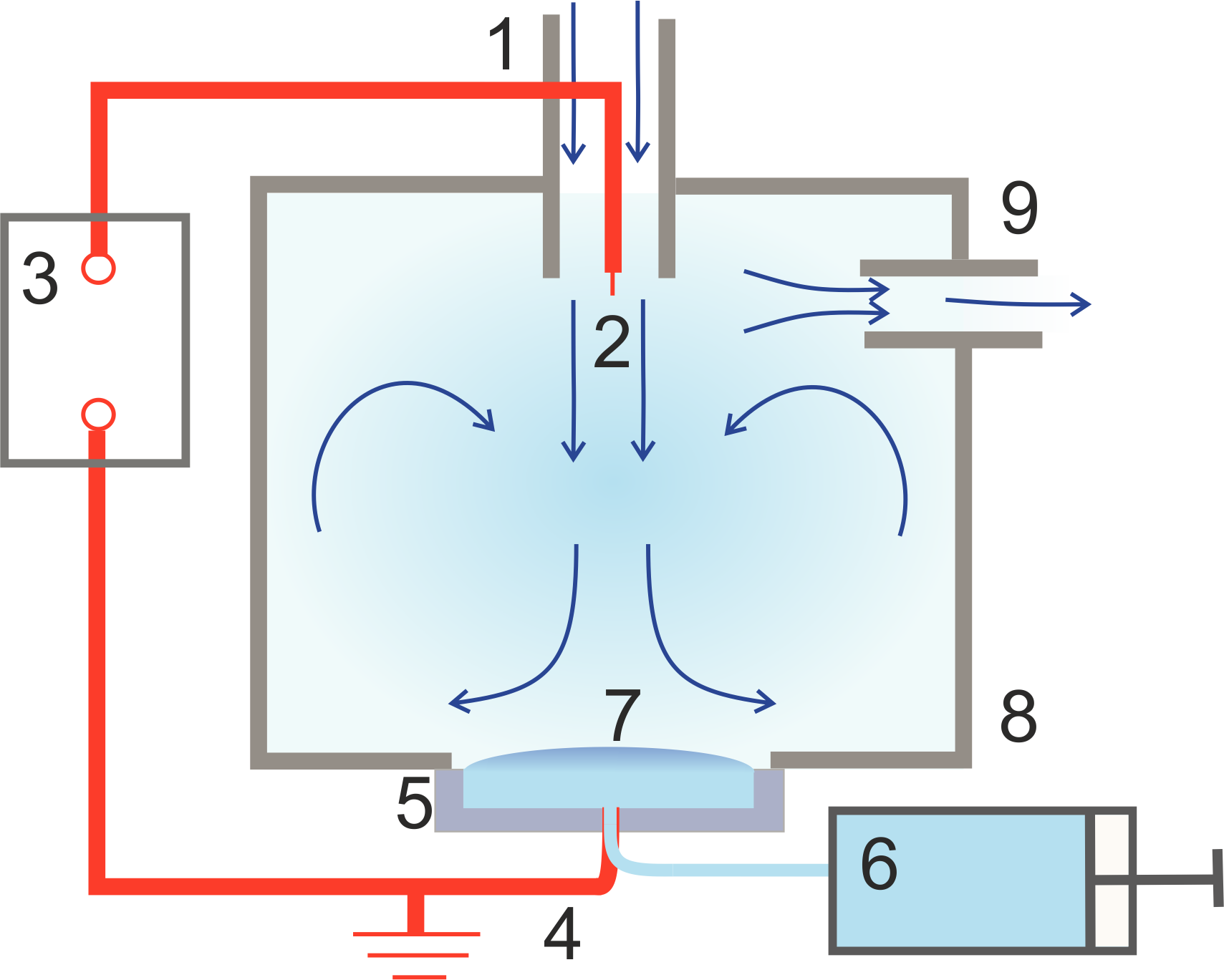}
\includegraphics[width=0.35\linewidth]{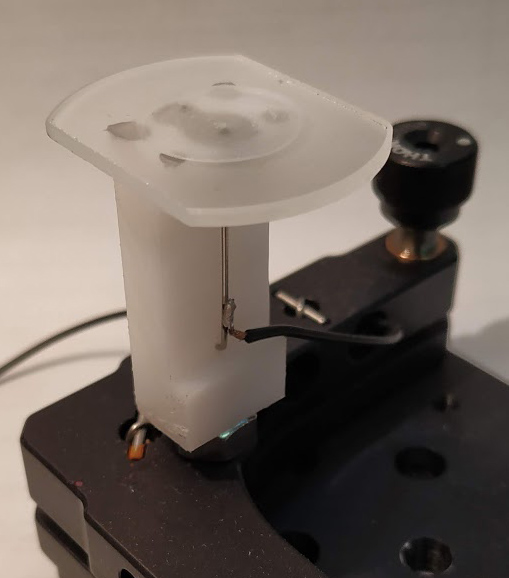}
\caption{Scheme of the ESAT system (left panel) and an overview of the cuvette (right panel). Numbers indicate parts of the setup: \textit{1} - silicone tube, \textit{2} - thin stainless steel needle, \textit{3} - high voltage power supply , \textit{4} - grounded collecting electrode,  \textit{5} - cylindrical glass cuvette, \textit{6} -  syringe, \textit{7} -  water surface, \textit{8} -  cylindrical plastic box, \textit{9} - hole for the air outlet.}
\label{fig:Cuvette}
\end{figure}

The schematic diagram of the ESAT system is shown in Fig.\ref{fig:Cuvette}. The exhaled air moves through a  silicone tube \textit{1} of 1.2~cm inner diameter. A thin stainless steel needle \textit{2} with 20~$\mu$m tip is installed along the centerline of the tube just before its outlet in such a way that the needle tip is 0.5~cm below the tube edge. The needle is connected to a negative electrode of the high voltage power supply \textit{3}, which results in inducing a corona discharge ionizing the air molecules near the needle tip. The aerosol droplets emerged from the tube collide with the ionized air molecules, thus becoming electrically charged, and then they are transported by the electrostatic force toward the grounded collecting electrode \textit{4}. The latter is placed at the bottom of the cylindrical glass cuvette \textit{5} of 1.6~cm in diameter and 0.05~cm in depth.  Before each experiment the cuvette was thoroughly cleaned and filled with high purity water  up to its edge. The electrode is a thin stainless tube (the outer diameter is 0.07~cm) connected to a syringe \textit{6} to vary the water volume in the cuvette (with accuracy 0.1~$\mu$L), which compensates the volume loss of the water due to evaporation. The droplets moving toward the collecting electrode coalesce with the water surface \textit{7}  and accumulate there in the form of an adsorbed film. Further, the surface-active properties of the film were investigated by the tensiometric method. The ESAT system was isolated from the environment by a cylindrical plastic box \textit{8}, which prevented contamination of a sample. The box has a hole \textit{9} for the air outlet.

%%%%%%%%%%%%%%%%%%%%
%     FIGURE 1     %
%%%%%%%%%%%%%%%%%%%%
%%%%%%%%%%%%%%%%%%%%

The efficiency of droplet trapping by the ESAT system may depend on the potential difference between the electrodes, interelectrode distance, number and geometry of the electrodes, and the air flow rate. To optimize the configuration of the ESAT system and to find the regime with maximal trapping efficiency, additional experiments with a model aerosol were performed. 

\subsection{Optimization of the ESAT system in the experiments with a model aerosol} 
\label{sec:ESAToptimization}

\begin{figure}[h]
\includegraphics[width=1\linewidth]{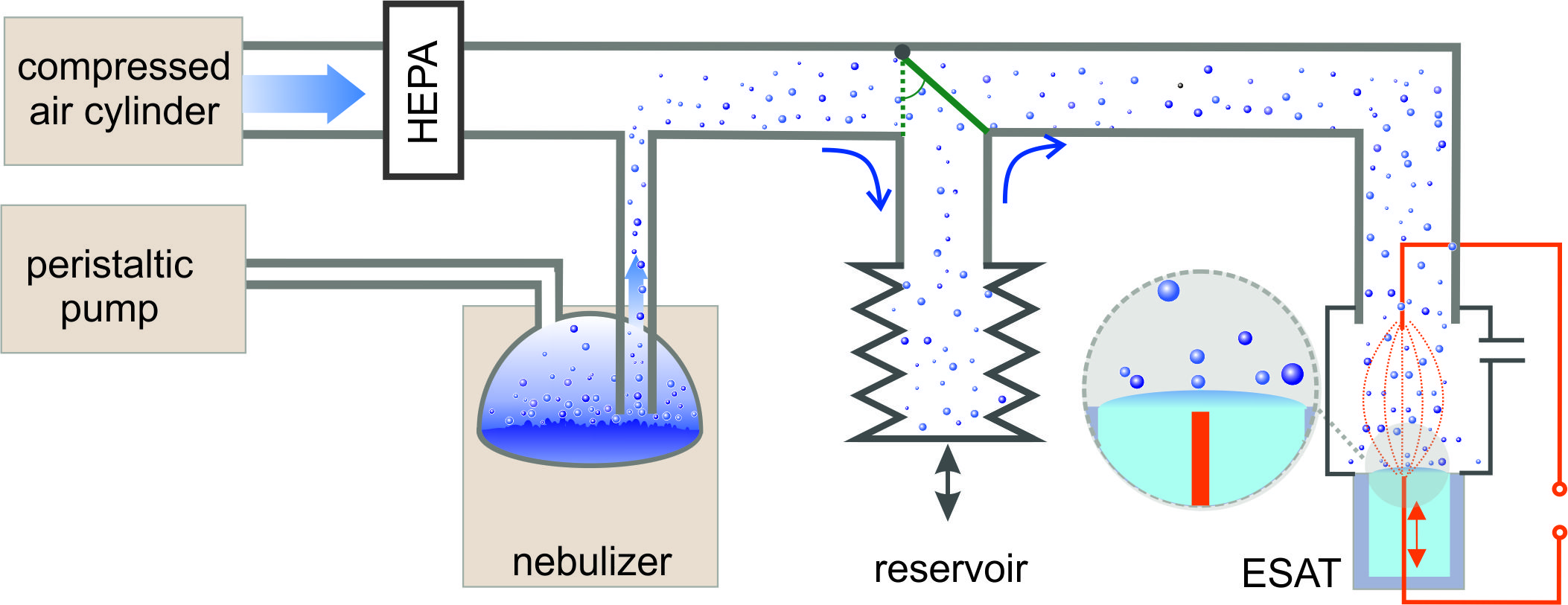}
\caption{Scheme of the experimental setup used for optimization of the ESAT parameters using the model aerosol.}
\label{fig:Inst1}
\end{figure}

The model aerosol was generated by the nebulizer (Omron-NE-U17, OMRON Corporation, Japan) which produced the droplets with a mass median diameter of 4.4~$\mu$m at a nebulization rate up to 3~mL/min. The peristaltic pump (see Fig.~\ref{fig:Inst1}) pushed the constant air flow through the nebulizer container. The concentrated air-droplet mixture, coming from the nebulizer container, was supplied through a T-joint into a silicone tube, where it was mixed with the air flux, forming thus the model aerosol, which entered the ESAT system. The constant air flow rate was provided by the compressed air cylinder connected to the silicone tube through a pressure reducer.  Such design allowed us to vary separately the aerosol concentration and its flow rate through the ESAT system by changing the nebulizing rate and the rate of the air flow from the compressed air cylinder. To prevent contamination of the model aerosol from the outside environment, the air, coming from the peristaltic pump and the compressed air cylinder, was infiltrated with HEPA filters. 

%%%%%%%%%%%%%%%%%%%%
%     FIGURE 2     %
%%%%%%%%%%%%%%%%%%%%
%%%%%%%%%%%%%%%%%%%%

A 10\%-by-mass aqueous solution of sodium chloride was used to produce a model aerosol. The saline solution allows us to perform accurate control over the amount of trapped droplets by measuring the conductivity of water in the cuvette. The cuvette, used in the experiments with the model aerosol, differs in a depth (being 50~mm deep) from that for the exhaled aerosol.

For conductivity measurements, the sensor of the conductometer (LR01V, WTW, Germany), having the measuring range 0.001-200~$\mu$S/cm, was placed in the water sample after each experiment. During the collecting procedure, the sensor was taken out and the grounded electrode was put inside the water through the bottom so that its tip was approximately 1~mm below the liquid surface.

As a measure of the effectiveness of the ESAT system, the ratio of the saline mass $m_{tr}$, trapped in the cuvette, and its total mass $m_{tot}$, delivered to the air flow from a nebulizer during the collecting procedure, was used

$$K=\frac{{{m}_{tr}}}{{{m}_{tot}}}\cdot 100\%$$ 

The value $m_{tot}$ was defined as $m_{tot}=Mt$, where $t$ is the experiment duration and $M$ is the saline mass rate which depends on both the nebulizer capacity and the flow rate provided by the peristaltic pump. The $M$ values were preliminary measured for each regime by means of weighing the nebulizer reservoir before and after a longtime experiment. The saline mass $m_{tr}$, trapped in the cuvette during the experiment, was calculated from the measured values of water conductivity obtained at the end of each experiment. 

The experimental procedure was as follows. The cuvette was filled with pure water and the conductivity was measured to check the purity of the trapping system. The grounded electrode was set into the water. The nebulizer, the peristaltic pump and the air delivery from the compressed air cylinder were switched on. After the aerosol filled completely the tube from T-joint to ESAT, the high voltage power supply was switched on. From this moment the experiment duration time was counted. Upon completion of the trapping procedure, the high voltage power supply was switched off, the aerosol delivery was stopped and the grounded electrode was taken out of the cuvette. After that the liquid in the cuvette was mixed by a magnetic stirrer for 2~minutes, and then its conductivity and temperature were measured. Each experiment was repeated 5 times to estimate an experimental error. During the experiments we had been measuring the aerosol trapping effectiveness depending on the system configuration, namely, on the distance $D$ between the tube edge and the liquid surface, the air flow rate $Q$, the saline mass flow $M$ produced by a nebulizer and the potential difference $U$ between the electrodes. The variations of the parameters were made relative to the so called base configuration which was as follow: $D=4.0~cm$, $Q=0.5~L/s$, $M=250~\mu g/s$, $U=5.0~kV$. 

First, we found that the droplet trapping is extremely inefficient without use of the electric field. In the base configuration and at $U=0~kV$, the effectiveness was only $K=(0.05\pm0.02)\%$. This means that the aerosol droplets, being extremely small in size, move with the air not deviating from streamlines. Under these conditions, the rare collisions with the water surface become possible only for those droplets which move along the streamlines lying very close to the interface. Creation of the potential difference between the electrodes equal to $5.0~kV$ raises the effectiveness to the order up to $0.5\%$. At $U=20~kV$, it becomes already $K=2.0\%$, which indicates a linear dependence of these parameters. It was not possible to use higher voltage due to the electrical breakdown occurrence. 

Approaching the tube edge toward the water surface also increases the amount of the trapped droplets. A decrease in the distance $D$ up to 2~cm increases the effectiveness twice. Further reduction of the distance results in strong perturbations of the water surface with the air jet which may lead to splashing the water out of the cuvette even at low air flow rates. The flow rate variation has the most profound effect on the efficiency. The reduction of this parameter from $0.5~L/s$ to $0.1~L/s$ increases $K$ from $0.5\%$ to $5.0\%$. The decrease of the saline mass flow by one order up to $M=25 \mu g/s$ had no effect on the effectiveness of the ESAT system. In our opinion, this fact offers the possibility of extending  the results obtained with the model aerosol to the case of the exhaled aerosol where the mass flow  is still four orders of magnitude lower. 

Thus, the optimal set of parameters obtained in the experiments with the model aerosol is as follows: $D=2.0~cm$, $Q=0.1~L/s$, $U=20.0~kV$. The effectiveness of the ESAT system was found to be near 30\% under these conditions, and it may increase still more at lower values of the air flow rate. However, the maintenance of such low exhalation rate is complicated for a subject and reduces significantly the exhaled particle concentration \cite{Johnson2009}. On the other hand, use of the flow rate of $1.0~L/s$, corresponding to normal breathing, leads to a considerable reduction of the trapping effectiveness. To handle this problem, an intermediate flexible reservoir was installed into the setup construction (see Fig.~\ref{fig:Inst1} and Fig.~\ref{fig:Inst}). 

A subject makes a normal breathing into this reservoir through automatic 2-way valve. When an exhalation is finished the exhaled air is pressed out with a slow flow rate into ESAT tract by applying a load to the reservoir. The use of the additional reservoir for intermediate storage has certain disadvantage. Some of the particles are inevitably lost due to settling on the walls of the reservoir, which decreases the effectiveness of the trapping process. Nevertheless, the gain in effectiveness caused by the reduction of the air flow rate turns out to be higher. The experiments with the model aerosol have shown that the use of reservoir reduces the effectiveness three times, whereas the decrease of the air flow rate from $1.0~L/s$ to $0.1~L/s$ increases the effectiveness twenty times. Finally, the following set of parameters was selected for further experiments with exhaled air: $D=2.0~cm$, $Q=0.1~L/s$ (here, $Q$ is the air flow rate provided taking into account the application of the intermediate reservoir), $U=20.0~kV$. Under these conditions, the effectiveness was about 10\%. 

\subsection{The setup for trapping exhaled particles.} 
\label{sec:Exhaled air}

\begin{figure}[h]
\includegraphics[width=1\linewidth]{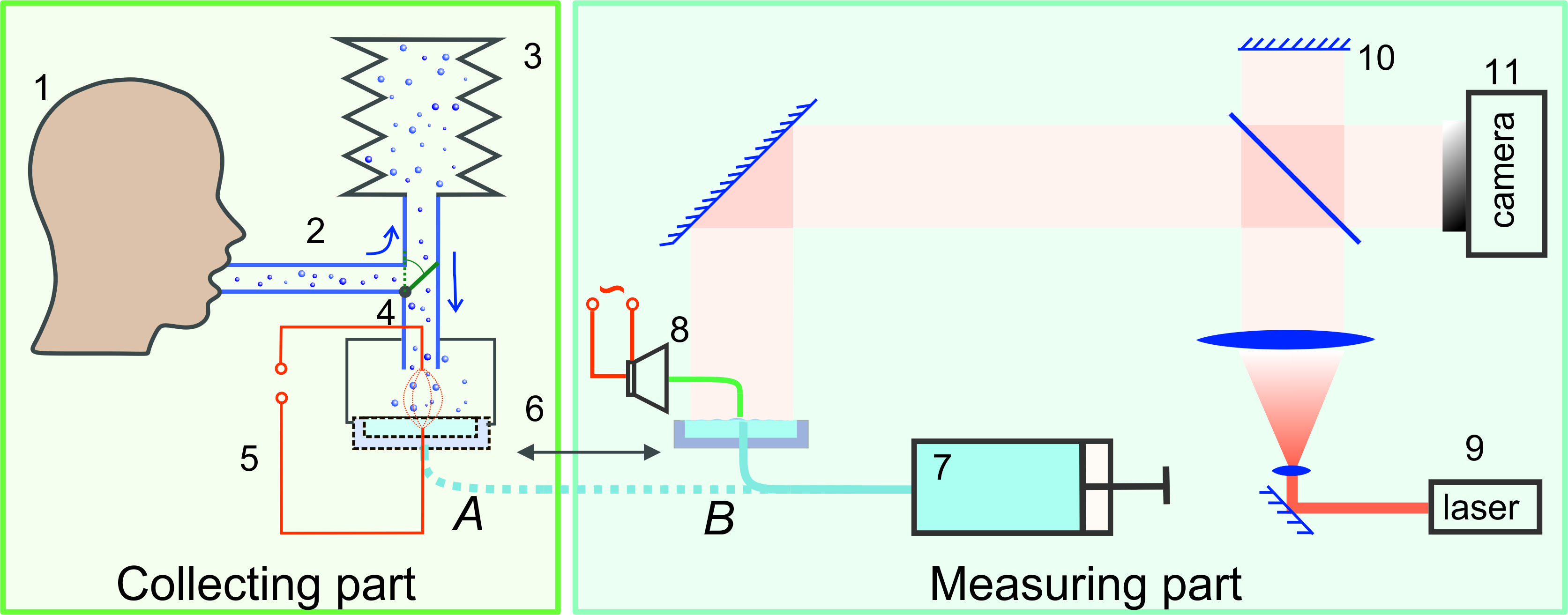} 
\caption{Scheme of a setup for collecting and analyzing the exhaled particles. The collecting module includes: \textit{1} - subject, \textit{2} - silicone tube, \textit{3} - reservoir, \textit{4} - automatic 2-way valve, \textit{5} - high voltage power supply, \textit{6} - cuvette. The measurement module includes: \textit{7} - syringe, \textit{8} - speaker,  \textit{9} - laser, \textit{10} - reference glass plate, \textit{11} - video camera. $A$ and $B$ are the positions of the cuvette during the collecting and measuring procedures, respectively. }
\label{fig:Inst}
\end{figure}

The setup for trapping native aerosol particles from exhaled air is presented in Fig.~\ref{fig:Inst}. The design of the collecting module is similar to that used in the experiments with the model aerosol. After the cuvette in the ESAT system (position $A$ in Fig.~\ref{fig:Inst}) was filled with water and the high voltage power supply was switched on, a subject \textit{1} made one full expiration at the flow rate approximately $1.0~L/s$ through the tube \textit{2} (see Fig.~\ref{fig:Inst}). To avoid any contamination of the inhaled air from the environment, the subject inhaled filtered, particle-free room air through a filter. The exhaled air came through the automatic 2-way valve \textit{4} into the intermediate storage reservoir \textit{3}. After the exhalation was finished, the air was pressed out from the reservoir \textit{3} into the ESAT tract with the slow flow rate of $0.1~L/s$. When the air completely came out from the reservoir, the subject made next expiration. 

%%%%%%%%%%%%%%%%%%%%
%     FIGURE 3     %
%%%%%%%%%%%%%%%%%%%%
%%%%%%%%%%%%%%%%%%%%%

After the required number of exhalations was made, the high voltage power supply \textit{5} was switched off, and the cuvette was shifted to position B (see Fig.~\ref{fig:Inst}) so that the water surface containing a collected material can be analyzed using a capillary waves method. 

\subsection{Capillary waves method} 
\label{sec:capwave}
We used the modified version of the capillary waves method which was described in detail in Ref.~\cite{Shmyrov2019}. The capillary wave was excited by periodical pressure pulsations in the gas phase near the interface. For this purpose, the acoustic wave generated by the miniature speaker \textit{8}, connected to the AC generator, was directed to the water surface through the thin steel tube ($0.08~cm$ in diameter and $2.5~cm$ length) used as a waveguide. The face of the tube was parallel to the liquid surface and situated at a distance of $0.01~cm$. The frequency range was from 0.5 to 2.5~kHz. 

Optical interferometry was used to register the interface relief. On the contrary to the optical scheme described in \cite{Shmyrov2019}, we applied the Michelson scheme which is more suitable in the considered case. The interference pattern resulting from the addition of two collimated laser beams reflected from the surface of the liquid and the reference glass plate \textit{10} was observed using a video camera \textit{11} (TXG50, Baumer, Germany). The view field of the camera is 2~cm, and the matrix resolution is 2500$\times$2000 pixels, which ensures spatial resolution of 10~$\mu$m/pixel. The reference plate and the unperturbed liquid interface form the angle of approximately $30'$, which allows us to observe the interface deformation in the interference fringes of constant thickness. This approach makes it possible to apply the spatial phase shifting technique for precise reconstruction of the interface relief. 

\subsection{Data processing} 
\label{sec:dataprocessing}

Each interferogram were processed in three steps. First, a 3D interface profile was reconstructed from the interferogram by means of the spatial phase shifting method using the IntelliWave (ESDI, USA) software. From the 3D profile, we filtered only deformations associated with small-scale capillary waves, as it was described in detail in \cite{Shmyrov2019}. The profile thus obtained was approximated by the damped cylindrical wave equation: 

\begin{equation}
z(\vec{r})=\frac{A}{\sqrt{|\vec{r}|}} e^{-\beta |\vec{r}|} \Re \left(e^{-i (\omega t - \vec{k}\vec{r})}\right)+z_0,
\end{equation} 

where $A$ is the wave amplitude, $\omega = 2 \pi \nu$ is the excitation frequency, $t$ is the time, $\vec{r}$ is the radius vector from the point of wave excitation, $\vec{k}$ is the wave vector, and $\beta$ is the attenuation coefficient. This model well describes a capillary wave away from the source (beyond about three wavelengths) with an accuracy of $0.05\%$. The central part of the wave that is close to the source was excluded from consideration. The surface tension $\sigma$ and the attenuation coefficient $\beta$, calculated as described in Ref.~\cite{Shmyrov2019}, were used to characterize the properties of the adsorbed film occurring on the water surface during the trapping of exhaled particles. 

\section{Results and discussion.}
\label{sec:res}

\begin{figure}[h]
\includegraphics[width=1\linewidth]{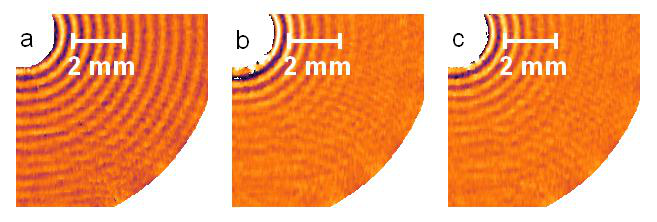}
\includegraphics[width=1\linewidth]{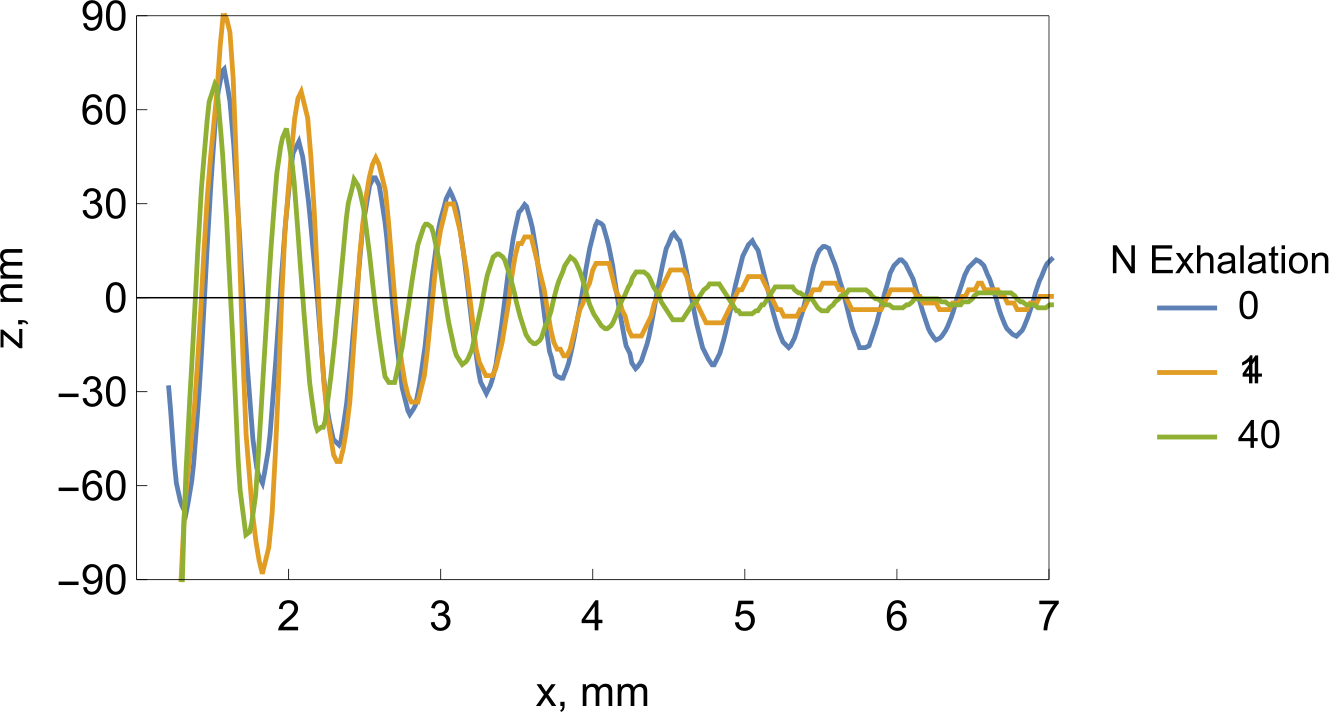}
\caption{3D profiles of the capillary waves observed at pure water (a) and the water surface obtained after 14 (b) and 40 (c) exhalations of the subject. (d) - vertical cross sections of the profiles (a), (b) and (c).} 
\label{fig:Images}
\end{figure}

Typical capillary wave profiles observed at the pure water surface (Fig.~\ref{fig:Images}~a) and at the water surface obtained after 14 (Fig.~\ref{fig:Images}~b) and 40 (Fig.~\ref{fig:Images}~c) exhalations of a subject are presented in Fig.~\ref{fig:Images}. The form of the wave changes as the number of exhalations increases. The changes become more evident from the analysis of the vertical cross sections of presented profiles (Fig.~\ref{fig:Images}~d). The reduction of the wavelength and, consequently, the surface tension, as far as an increase in the attenuation coefficient, clearly indicate the accumulation of surface-active substances on the water surface. It is evident that these substances, constituent to the PS complex, accumulate on the water surface during the trapping process. 

%%%%%%%%%%%%%%%%%%%%
%     FIGURE 4     %
%%%%%%%%%%%%%%%%%%%%
%%%%%%%%%%%%%%%%%%%%

To make sure that the collected material comes to the water surface from the exhaled air rather than from the environment, we performed some additional experiments. In the first experiment, we used the filtered air coming to the ESAT system instead of the exhaled air. The water surface in the cuvette remained clean even after 20 min of the experiment. In the second test, we also used the filtered air instead of the exhaled air, but additionally we removed the plastic box covering the ESAT system. A noticeable reduction in the surface tension was found already in 3~minutes, which indicates pollution of the water surface with some inclusions trapped from the surrounding air by an electrostatic field. These tests show that the covering plastic box safely protects the ESAT system from outward contamination. 

\begin{figure}[h]
\includegraphics[width=1\linewidth]{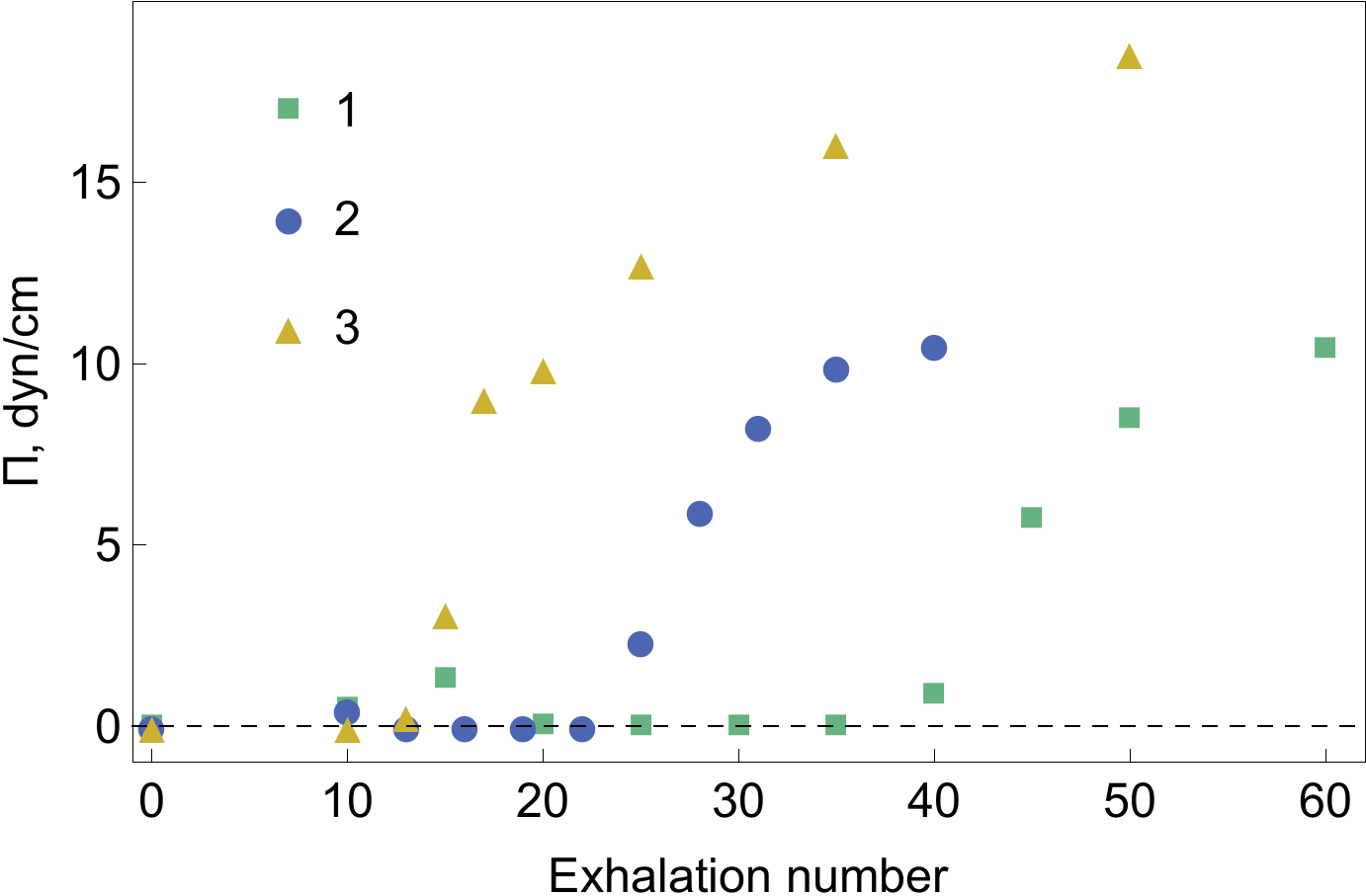} 
\caption{Variation of the surface pressure $\Pi$  with the number of exhalations for different voltage on the electrodes $1$ - 8~kV, $2$ - 10~kV, $3$ - 15~kV.}
\label{fig:PIVolt}
\end{figure}

Figure~\ref{fig:PIVolt} illustrates the dependence of the surface pressure $\Pi$ on the number of exhalations obtained at three different values of potential difference. The surface pressure was defined as $\Pi=\sigma_0-\sigma$ (here $\sigma_0$ and $\sigma$ is the surface tension of a clean water surface and the water surface containing an adsorbed film, respectively); it is equal to zero at clean interface. 

%%%%%%%%%%%%%%%%%%%%
%     FIGURE 5     %
%%%%%%%%%%%%%%%%%%%%
%%%%%%%%%%%%%%%%%%%%

The growth of the surface pressure is clear evidence of the accumulation of the surface-active substances contained in the exhaled particles on the free water surface. Under the assumption that approximately the same amount of the native material comes with every exhalation, the plot in Fig.~\ref{fig:PIVolt} may be considered as the dependence of the surface pressure $\Pi$ on the surface concentration $\Gamma$ of the surfactant adsorbed at the water surface, which is called the surface pressure isotherm. 

The form of the curves presented in Fig.~\ref{fig:PIVolt} is typical for the isotherm of any surfactant. At the initial segment of the curve, when the surface concentration is low, the surfactant molecules are in gaseous phase state and, therefore, interact weakly. As the surface concentration increases, the surface pressure in such rarefied layer grows slowly and only slightly differs from zero. At higher surface concentration, when the molecules begin to interact, the film is in a liquid-expanded phase state, which is characterized by a higher growth rate of surface pressure as the surfactant content at the interface increases. The form of the $\Pi(\Gamma)$ curve is a unique characteristic of any surfactant. Three curves in Fig.~\ref{fig:PIVolt} plotted for different values of the potential difference  between the electrodes in the ESAT system are similar because they reflect accumulation of the same PS but at different rate. 

%%%%%%%%%%%%%%%%%%%%
%     FIGURE 6     %
%%%%%%%%%%%%%%%%%%%%
\begin{figure}[h]
\includegraphics[width=1\linewidth]{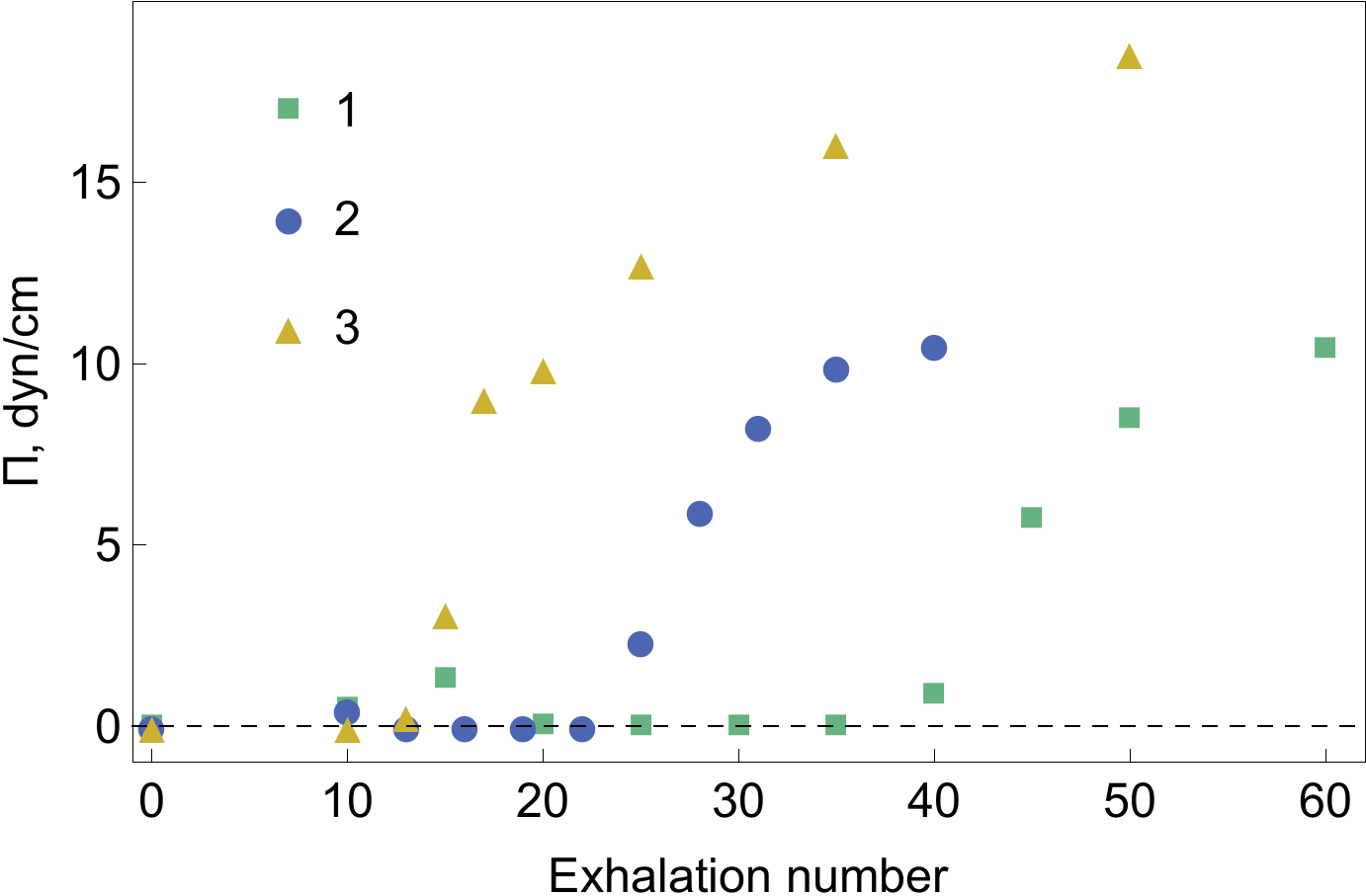} 
\caption{Variation of the surface pressure $\Pi$  with a number of exhalations.}
\label{fig:PIvsEx}
\end{figure}
%%%%%%%%%%%%%%%%%%%%

In contrast to the experiments with the model aerosol, we have found the nonmonotonic dependence of the efficiency of the trapping process on the voltage applied. At values below 15~kV, the surface pressure, measured on the water surface after a fixed number of exhalations, grows almost linearly as the potential difference increases (see Fig.~\ref{fig:PIVolt}). However, further increase of the voltage up to 17~kV leads to unexpected result, namely, we found that the surface pressure remains equal to zero after any quantity of exhalations. Moreover, the native material, collected ,e.g., at 15~kV, disappears from the interface in a few seconds after we increase the potential difference to 17~kV. In our opinion, this phenomenon can be caused by ozone production in the corona discharge at high enough voltage. The characteristic smell of this substance begins to be felt at voltages above 15~kV. Being strong oxidizer, ozone breaks the organic molecules that are constituents of the PS complex, cleaning thus the water surface. Taking this fact into account, the maximal value of the potential difference, used in the further experiments, did not exceed 13~kV. 

To check the possibility of producing the PS saturated monolayer and to estimate an absolute amount of the collecting material, we examined the $\Pi$ variations during the long-term collection. To this end, we made measurements after every ten-twenty exhalations. The results of this experiment are presented in Fig.~\ref{fig:PIvsEx}. It is seen that the $\Pi$ does not reach saturation, continuing to grow according to logarithmic law (see inset in Fig.~\ref{fig:PIvsEx}) even after three hundred exhalations.

For most surfactants, the surface concentration value at which the phase transition occurs in the film from the gas to the liquid-expanded state is ${\Gamma }/{{{\Gamma }_{e}}}\;=\left( 0.3-0.5 \right)$, where ${{\Gamma }_{e}}$ is the surface concentration in a saturated film. Taking into account the fact that in our experiments the phase transition in the adsorbed film is observed near the fifteenth exhalation (at $15~kV$), one can expect the saturation starting with approximately 40-60~th exhalations. The results presented in Fig.~\ref{fig:PIvsEx} show the surface pressure increases slower compared to the expected value, which indicates that the ESAT system effectiveness decreases during the collection procedure. On the contrary, this was not observed in the experiments with the model aerosol, in which the efficiency remained constant over time. It is obvious that this should be somehow related to the properties of the material being collected. 

In our opinion, the most probable cause of a reduction in the trapping efficiency is a deterioration of coalescence conditions due to the formation of an adsorbed film. The PS which covers both particle and water surfaces prevents their approaching at the distance at which coalescence occurs. Moreover, this effect reinforces as the surfactant accumulates. A similar situation is known to arise in emulsions, the stability of which against aggregation is achieved by forming a surfactant film on the droplet surface. 

For the correct analysis of the PS surface properties, it is necessary to investigate the relationship between the surface pressure and the surfactant surface concentration $\Gamma$, which should be obtained from the analysis of the dependence of $\Pi$ on the number of exhalations (see, e.g., Fig.~\ref{fig:PIvsEx}).  However, as the quantity of the collected material is nonlinearly related to the amount of exhaled air passed through the system, this procedure is not evident. In addition, it is not convenient for the subject to be interrupted for periodic measurements during the collection process. Therefore, we changed the sequence of actions to solve the above problems. 

Initially, the material is accumulated on the water surface during the sequential 50 exhalations of the subject. Further, the work is done by an operator in the absence of a subject. The operator inserts the cuvette into the part of the experimental setup where the capillary wave technique is realized and then measures the wavelength and the attenuation coefficient of the wave at 3-5 frequencies. After that, the operator removes some amount of an adsorbed substance from the surface using a sampler. The sampler is a thin-walled metal tube with an inner diameter of 2.0~mm, the inner and outer surfaces of which are coated with an anti-wetting agent. A small lenticular drop of liquid coated with a surfactant remains on the tip of the sampler after each touching, reducing thereby the surface concentration of the surfactant in the cuvette. In our experiments, the ratio of the surface area of the droplet to be removed and the water surface in the cell was such that half of the adsorbed material was removed after 25 touches of the sampler. The wave profiles were measured after each change in the surface concentration. 

\begin{figure}[h]
\includegraphics[width=1\linewidth]{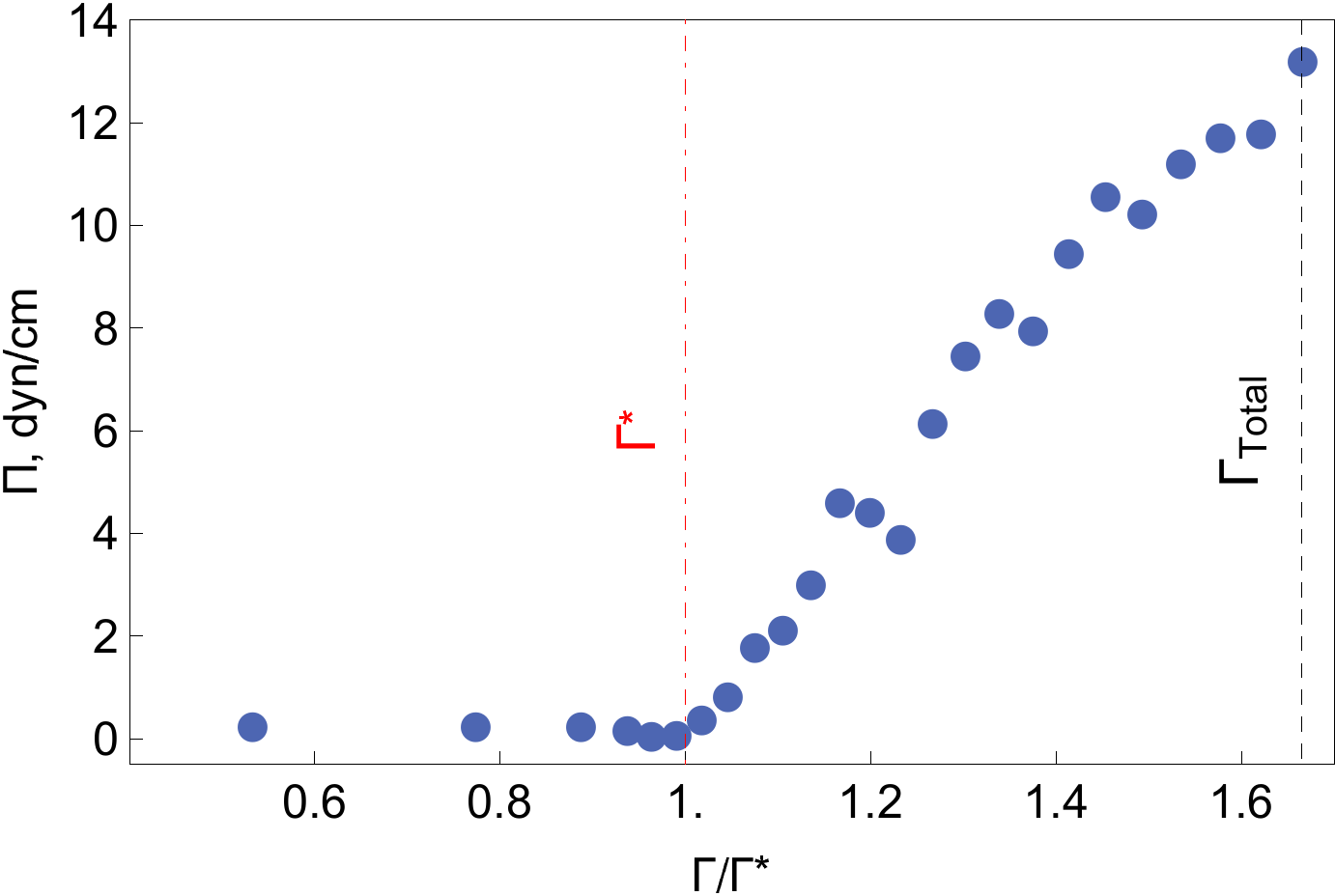}
\includegraphics[width=1\linewidth]{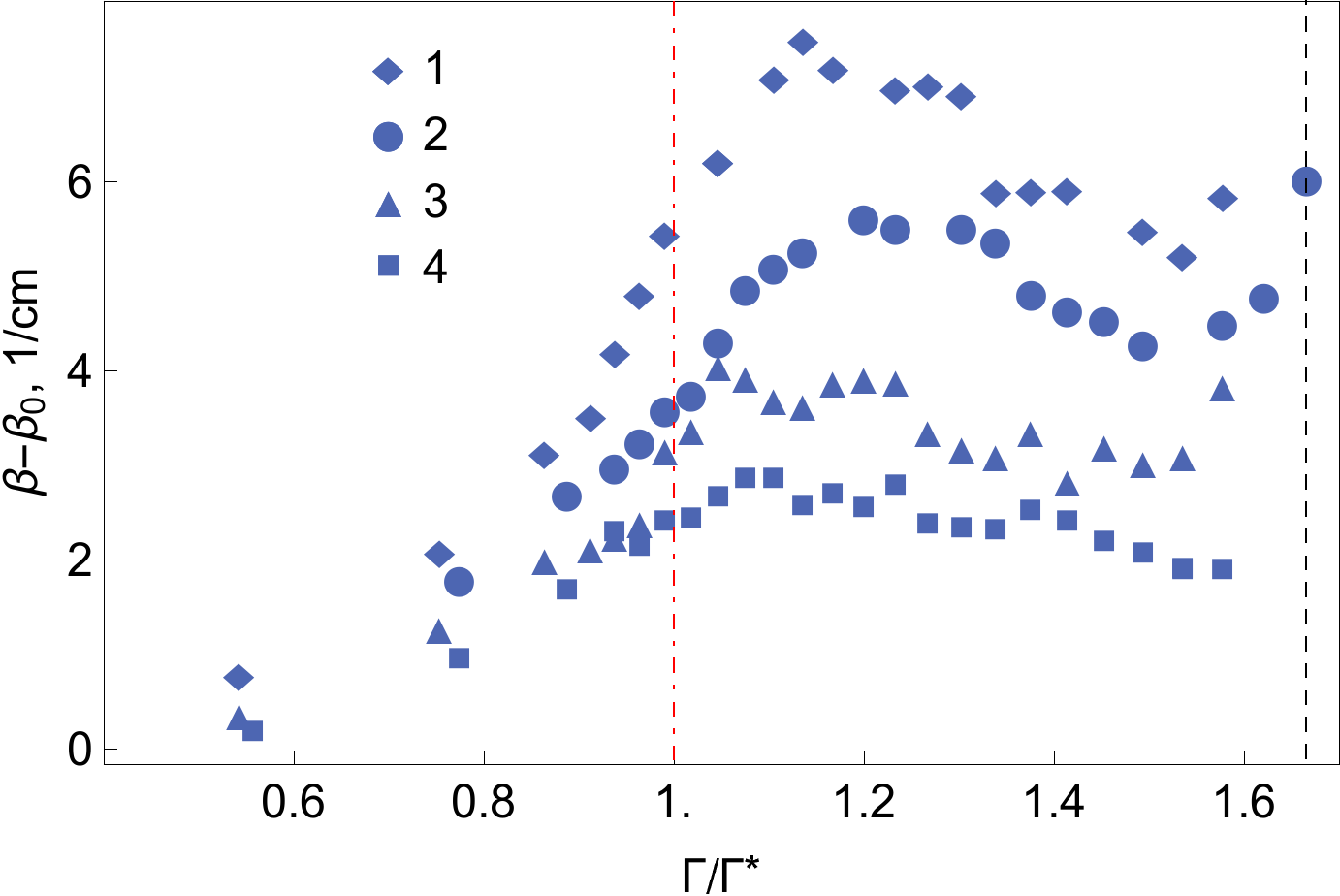} 
\caption{Upper panel - dependence of the surface pressure $\Pi$ on the surfactant concentration $\Gamma/\Gamma^*$. Lower panel - dependence of $\beta-\beta_0$ on $\Gamma/\Gamma^*$ for different frequencies of wave excitation \textit{1} - 2.5 kHz, \textit{2} - 2 kHz, \textit{3} - 1 kHz,\textit{4} - 0.7 kHz. Red dash-dotted line indicates $\Gamma^*$ at which a phase transition in the surfactant film from the gaseous phase state to the liquid-expanded state was observed. Black dashed line denotes $\Gamma_{total}$ - the surface concentration achieved after the total number of subject's exhalations. }
\label{fig:PIandBeta}
\end{figure}

The surface pressure calculated using the wavelength measurements and the attenuation coefficient of the capillary wave are presented in Fig.~\ref{fig:PIandBeta}. Note that both values were determined for each surface concentration obtained by the method described above. The surface concentration was measured in relative units $\Gamma/{\Gamma^*}$, being normalized to the value of the surface concentration $\Gamma^*$ at which a phase transition in the surfactant film from the gaseous phase state to the liquid-expanded state was observed. The attenuation coefficient is measured as $\beta - {{\beta} _ {0}}$, where $ {{\beta} _ {0}} $ is the attenuation coefficient taken at $\Gamma =0$, i.e. on the free surface of water, and occurred due to the volumetric viscosity of water. 

%%%%%%%%%%%%%%%%%%%%
%     FIGURE 7     %
%%%%%%%%%%%%%%%%%%%%
%%%%%%%%%%%%%%%%%%%%

A comparison of the two dependencies shows that the attenuation coefficient is more sensitive to the presence of the collected material on the surface. This parameter begins to increase noticeably already at small values of the surface concentration, whereas the surface pressure begins to increase only after the transition of the adsorbed layer into a liquid-expanded state. It is worth noting that, in  contrast to the $\Pi(\Gamma)$ dependence, the $\beta(\Gamma)$ one is nonmonotic and exhibits a maximum associated with resonance effects. The excitation of a capillary wave on the water surface containing a surfactant is accompanied by the appearance of a dilatation wave associated with stretching and compression of the surfactant film. This additional dissipative mechanism amplifies wave attenuation. It is known that the local maximum of the attenuation coefficient of the capillary wave is observed when its frequency coincides with the natural frequency of the dilatation wave \cite{vogel1989resonance}. Varying the capillary wave frequency (see Fig.~\ref{fig:PIandBeta}) causes the local maximum of this dependence to shift because the natural frequency of the dilatation wave is the function of the surface surfactant concentration.

\section{Conclusion} 
\label{sec:concl}
  
In this paper we demonstrate the possibility of using electrostatic forces to trap aerosol particles containing in the exhaled air on a gas-liquid surface. The PS components containing in the particles accumulate on the liquid surface in the form of an adsorbed film, the surface-active and rheological properties of which can be investigated by the tensiometric method. The high efficiency of the ESAT system proposed in this paper, as well as the small liquid surface area required for the implementation of the modified method of capillary waves, make it possible to collect native material in an amount sufficient for tensiometric study during only 30-50 exhalations of the subject. The modified capillary waves method allows studying the dependence of the surface pressure and attenuation coefficient on the relative surface concentration of PS. The shape of both curves depends on the composition of the adsorbed layer and is its unique characteristic. 

PS is a complex of a few surface-active substances and each component makes its own contribution to the surface properties of the adsorbed layer \cite{longo1993function,schurch2001surface,Possmayer2001,ma2006real,bykov2019dynamic}. Variations in the components ratio or their quantity caused by the presence of a pulmonary disease are able to vary the shape of both $\Pi(\Gamma)$ and $\beta(\Gamma)$ dependencies, which can be used as the markers of a PS system dysfunction. For comparative analysis of PS samples in healthy subjects and patients with pulmonary diseases, the following quantitative characteristics can be utilized: 

1. $\Gamma_{total}$ which indicates the surface concentration (expressed in relative units) achieved after completion of the total number of exhalations and can be applied to assess the PS concentration in exhaled air. 

2. $d\Pi/d\Gamma$ which is measured at $\Gamma >1$, i.e., on the part of the $\Pi \left( \Gamma  \right)$ curve where the adsorbed film is in the liquid-expanded phase state. This value reflects the surface activity of PS, i.e., it shows how quickly the surface tension decreases with increasing surface concentration. 

3. $\beta_{max}$ and $\Gamma_{max}$ which are the coordinates of the extremum on the $\beta\left(\Gamma\right)$ curve. 

4. $\beta\left(\Gamma=1\right)$ which denotes the value of the attenuation coefficient at the phase transition point. 

5. $d\Gamma_{max}/d\nu$ and $d\beta _{max}/d\nu$ which reflect the shift of the extremum position with changing the frequency of the wave excitation. 

The key feature of the proposed facility is the ability of point of care testing PS functional state. The sample of PS can be nonivasively collected and immediately examined. In some groups of subjects, such elder people, children, patients with distress syndrome this method can be the only alternative to collect lung native material. Moreover, there is a possibility to connect the ESAT system to a artificial lungs ventilation machine to control PS state even in intensive care. Both modules of the complex, ESAT and measurment part, can be fully automatized.

%%%%%%%%%%%%%%%%%%%%%%%%%%%%%%%%%%%%%%
% ====== REFERENCE SECTION===========%
%%%%%%%%%%%%%%%%%%%%%%%%%%%%%%%%%%%%%%


\begin{thebibliography}{10}

	\bibitem{Haslbeck2010}
	K.~Haslbeck, K.~Schwarz, J.~M. Hohlfeld, J.~R. Seume, and W.~Koch, ``Submicron
	droplet formation in the human lung,'' \emph{Journal of Aerosol Science},
	vol.~41, no.~5, pp. 429--438, 2010.
	
	\bibitem{Johnson2009}
	G.~R. Johnson and L.~Morawska, ``{The Mechanism of Breath Aerosol Formation},''
	\emph{{Journal of Aerosol Medicine and Pulmonary Drug Delivery}}, vol.~{22},
	no.~{3}, pp. {229--237}, {2009}.
	
	\bibitem{almstrand2010effect}
	A.-C. Almstrand, B.~Bake, E.~Ljungstr{\"o}m, P.~Larsson, A.~Bredberg,
	E.~Mirgorodskaya, and A.-C. Olin, ``Effect of airway opening on production of
	exhaled particles,'' \emph{Journal of applied physiology}, vol. 108, no.~3,
	pp. 584--588, 2010.
	
	\bibitem{Grotberg2011}
	J.~B. Grotberg, ``{Respiratory fluid mechanics},'' \emph{{Physics of Fluids}},
	vol.~{23}, no.~{2}, p. {021301}, {2011}.
	
	\bibitem{Holmgren2010}
	H.~Holmgren, E.~Ljungstrom, A.-C. Almstrand, B.~Bake, and A.-C. Olin, ``Size
	distribution of exhaled particles in the range from 0.01 to 2.0 $\mu$m,''
	\emph{Journal of Aerosol Science}, vol.~41, no.~5, pp. 439--446, 2010.
	
	\bibitem{Fairchild1987}
	C.~Fairchild and J.~Stampfer, ``Particle concentration in exhaled breath,''
	\emph{American Industrial Hygiene Association Journal}, vol.~48, no.~11, pp.
	948--949, 1987.
	
	\bibitem{Papineni1997}
	R.~Papineni and F.~Rosenthal, ``The size distribution of droplets in the
	exhaled breath of healthy human subjects,'' \emph{Journal of Aerosol
		Medicine: Deposition, Clearance, and Effects in the Lung}, vol.~10, no.~2,
	pp. 105--116, 1997.
	
	\bibitem{morawska2009size}
	L.~Morawska, G.~Johnson, Z.~Ristovski, M.~Hargreaves, K.~Mengersen, S.~Corbett,
	C.~Chao, Y.~Li, and D.~Katoshevski, ``Size distribution and sites of origin
	of droplets expelled from the human respiratory tract during expiratory
	activities,'' \emph{Journal of Aerosol Science}, vol.~40, no.~3, pp.
	256--269, 2009.
	
	\bibitem{Schwarz2010}
	K.~Schwarz, H.~Biller, H.~Windt, W.~Koch, and J.~M. Hohlfeld,
	``Characterization of exhaled particles from the healthy human lungвЂ”a
	systematic analysis in relation to pulmonary function variables,''
	\emph{Journal of aerosol medicine and pulmonary drug delivery}, vol.~23,
	no.~6, pp. 371--379, 2010.
	
	\bibitem{johnson2011modality}
	G.~Johnson, L.~Morawska, Z.~Ristovski, M.~Hargreaves, K.~Mengersen, C.~Chao,
	M.~Wan, Y.~Li, X.~Xie, D.~Katoshevski, \emph{et~al.}, ``Modality of human
	expired aerosol size distributions,'' \emph{Journal of Aerosol Science},
	vol.~42, no.~12, pp. 839--851, 2011.
	
	\bibitem{bake2017exhaled}
	B.~Bake, E.~Ljungstr{\"o}m, A.~Claesson, H.~K. Carlsen, M.~Holm, and A.-C.
	Olin, ``Exhaled particles after a standardized breathing maneuver,''
	\emph{Journal of aerosol medicine and pulmonary drug delivery}, vol.~30,
	no.~4, pp. 267--273, 2017.
	
	\bibitem{Olin2013}
	A.-C. Olin, ``{Particles in Exhaled Air-A Novel Method of Sampling
		Non-Volatiles in Exhaled Air},'' in \emph{{Volatile Biomarkers: Non-invasive
			Diagnosis in Physiology and Medicine}}, {Amann, A and Smith, D}, Ed.\hskip
	1em plus 0.5em minus 0.4em\relax Elsevier Science, {2013}, pp. {383--391}.
	
	\bibitem{Tinglev2016}
	A.~D. Tinglev, S.~Ullah, G.~Ljungkvist, E.~Viklund, A.-C. Olin, and O.~Beck,
	``{Characterization of exhaled breath particles collected by an electret
		filter technique},'' \emph{{Journal of Breath Research}}, vol.~{10}, no.~{2},
	p. {026001}, {2016}.
	
	\bibitem{Ullah2015}
	S.~Ullah, S.~Sandqvist, and O.~Beck, ``{Measurement of Lung
		Phosphatidylcholines in Exhaled Breath Particles by a Convenient Collection
		Procedure},'' \emph{{Analitical Chemistry}}, vol.~{87}, no.~{22}, pp.
	{11\,553--11\,560}, {NOV 17} {2015}.
	
	\bibitem{schwarz2015characterization}
	K.~Schwarz, H.~Biller, H.~Windt, W.~Koch, and J.~M. Hohlfeld,
	``Characterization of exhaled particles from the human lungs in airway
	obstruction,'' \emph{Journal of aerosol medicine and pulmonary drug
		delivery}, vol.~28, no.~1, pp. 52--58, 2015.
	
	\bibitem{Almstrand2009}
	A.-C. Almstrand, E.~Ljungstr{\"o}М€m, J.~Lausmaa, B.~Bake, P.~Sj{\"o}М€vall,
	and A.-C. Olin, ``Airway monitoring by collection and mass spectrometric
	analysis of exhaled particles,'' \emph{Analytical chemistry}, vol.~81, no.~2,
	pp. 662--668, 2009.
	
	\bibitem{Wurie2016}
	F.~B. Wurie, S.~D. Lawn, H.~Booth, P.~Sonnenberg, and A.~C. Hayward,
	``{Bioaerosol production by patients with tuberculosis during normal tidal
		breathing: implications for transmission risk},'' \emph{{Thorax}}, vol.~{71},
	no.~{6}, pp. {549--554}, {2016}.
	
	\bibitem{horvath2003exhaled}
	I.~Horv{\'a}th, ``Exhaled breath condensate contains more than only
	volatiles,'' \emph{European Respiratory Journal}, vol.~22, no.~1, pp.
	187--188, 2003.
	
	\bibitem{kazakov2000dynamic}
	V.~N. Kazakov, O.~V. Sinyachenko, V.~B. Fainerman, U.~Pison, and R.~Miller,
	\emph{Dynamic Surface Tensiometry in Medicine}, ser. Studies in Interface
	Science.\hskip 1em plus 0.5em minus 0.4em\relax Elsevier, 2000.
	
	\bibitem{baron2001aerosol}
	P.~A. Baron and K.~Willeke, \emph{Aerosol measurement: principles, techniques,
		and applications. 2nd ed.}\hskip 1em plus 0.5em minus 0.4em\relax
	Wiley-Interscience, New York, Nov 2001.
	
	\bibitem{notter2000lung}
	R.~Notter, \emph{Lung Surfactants: Basic Science and Clinical Applications},
	ser. Lung Biology in Health and Disease.\hskip 1em plus 0.5em minus
	0.4em\relax CRC Press, 2000.
	
	\bibitem{longo1993function}
	M.~Longo, A.~Bisagno, J.~Zasadzinski, R.~Bruni, and A.~Waring, ``A function of
	lung surfactant protein sp-b,'' \emph{Science}, vol. 261, no. 5120, pp.
	453--456, 1993.
	
	\bibitem{schurch2001surface}
	S.~Sch{\"u}rch, H.~Bachofen, and F.~Possmayer, ``Surface activity in situ, in
	vivo, and in the captive bubble surfactometer,'' \emph{Comparative
		Biochemistry and Physiology Part A: Molecular \& Integrative Physiology},
	vol. 129, no.~1, pp. 195--207, 2001.
	
	\bibitem{Possmayer2001}
	F.~Possmayer, K.~Nag, K.~Rodriguez, R.~Qanbar, and S.~Sch{\"u}rch, ``Surface
	activity in vitro: role of surfactant proteins,'' \emph{Comparative
		Biochemistry and Physiology Part A: Molecular \& Integrative Physiology},
	vol. 129, no.~1, pp. 209--220, 2001.
	
	\bibitem{ma2006real}
	G.~Ma and H.~C. Allen, ``Real-time investigation of lung surfactant respreading
	with surface vibrational spectroscopy,'' \emph{Langmuir}, vol.~22, no.~26,
	pp. 11\,267--11\,274, 2006.
	
	\bibitem{bykov2019dynamic}
	A.~Bykov, G.~Loglio, R.~Miller, O.~Milyaeva, A.~Michailov, and B.~Noskov,
	``Dynamic properties and relaxation processes in surface layer of pulmonary
	surfactant solutions,'' \emph{Colloids and Surfaces A: Physicochemical and
		Engineering Aspects}, vol. 573, pp. 14--21, 2019.
	
	\bibitem{Hohlfeld2002}
	J.~Hohlfeld, ``{The role of surfactant in asthma},'' \emph{{Respiratory
			Research}}, vol.~{3}, no.~{1}, {2001}.
	
	\bibitem{Baritussio2004}
	A.~Baritussio, ``{Lung surfactant, asthma, and allergens - A story in
		evolution},'' \emph{{American Journal of Respiratory and Critical Care
			Medicine}}, vol. {169}, no.~{5}, pp. {550--551}, {2004}.
	
	\bibitem{Wright2001}
	T.~Wright, R.~Notter, Z.~Wang, A.~Harmsen, and F.~Gigliotti, ``{Pulmonary
		inflammation disrupts surfactant function during Pneumocystis carinii
		pneumonia},'' \emph{{Infection and Immunity}}, vol.~{69}, no.~{2}, pp.
	{758--764}, {2001}.
	
	\bibitem{Schwab2009}
	U.~Schwab, K.~H. Rohde, Z.~Wang, P.~R. Chess, R.~H. Notter, and D.~G. Russell,
	``{Transcriptional responses of Mycobacterium tuberculosis to lung
		surfactant},'' \emph{{Microbial Pathogenesis}}, vol.~{46}, no.~{4}, pp.
	{185--193}, {APR} {2009}.
	
	\bibitem{Raghavendran2011}
	K.~Raghavendran, D.~Willson, and R.~N. Notter, ``{Surfactant Therapy for Acute
		Lung Injury and Acute Respiratory Distress Syndrome},'' \emph{{Critical Care
			Clinics}}, vol.~{27}, no.~{3}, pp. {525--559}, {2011}.
	
	\bibitem{Chroneos2009}
	Z.~C. Chroneos, K.~Midde, Z.~Sever-Chroneos, and C.~Jagannath, ``Pulmonary
	surfactant and tuberculosis,'' \emph{Tuberculosis}, vol.~89, pp. 10--14,
	2009.
	
	\bibitem{Chimote2005}
	G.~Chimote and R.~Banerjee, ``Lung surfactant dysfunction in tuberculosis:
	Effect of mycobacterial tubercular lipids on dipalmitoylphosphatidylcholine
	surface activity,'' \emph{Colloids and Surfaces B: Biointerfaces}, vol.~45,
	no. 3-4, pp. 215--223, 2005.
	
	\bibitem{Hasegawa2003}
	T.~Hasegawa and R.~M. Leblanc, ``Aggregation properties of mycolic acid
	molecules in monolayer films: a comparative study of compounds from various
	acid-fast bacterial species,'' \emph{Biochimica et Biophysica Acta (BBA) -
		Biomembranes}, vol. 1617, no. 1-2, pp. 89 -- 95, 2003.
	
	\bibitem{Wang2008}
	Z.~Wang \emph{et~al.}, ``Peripheral cell wall lipids of mycobacterium
	tuberculosis are inhibitory to surfactant function,'' \emph{Tuberculosis},
	vol.~88, no.~3, pp. 178 -- 186, 2008.
	
	\bibitem{PARDON2015}
	G.~Pardon, L.~Ladhani, N.~SandstrР“В¶m, M.~Ettori, G.~Lobov, and W.~van~der
	Wijngaart, ``Aerosol sampling using an electrostatic precipitator integrated
	with a microfluidic interface,'' \emph{Sensors and Actuators B: Chemical},
	vol. 212, pp. 344 -- 352, 2015.
	
	\bibitem{Morozov2017}
	V.~N. Morozov and A.~Y. Mikheev, ``A collection system for dry solid residues
	from exhaled breath for analysis via atomic force microscopy,'' \emph{Journal
		of Breath Research}, vol.~11, no.~1, p. 016006, 2017.
	
	\bibitem{Shmyrov2019}
	A.~Shmyrov, A.~Mizev, A.~Shmyrova, and I.~Mizeva, ``Capillary wave method: An
	alternative approach to wave excitation and to wave profile reconstruction,''
	\emph{Physics of Fluids}, vol.~31, no.~1, p. 012101, 2019.
	
	\bibitem{vogel1989resonance}
	V.~Vogel and D.~Moebius, ``Resonance of transverse capillary and longitudinal
	waves as a tool for monolayer investigations at the air-water interface,''
	\emph{Langmuir}, vol.~5, no.~1, pp. 129--133, 1989.
	
\end{thebibliography}
\end{document}